\documentclass[iop,apj,tighten]{emulateapj}
\usepackage{amsmath,amstext}
\usepackage{apjfonts}
\usepackage[breaklinks,colorlinks,citecolor=blue,linkcolor=magenta]{hyperref} 
\usepackage{float}
\usepackage{footmisc}
\usepackage{multirow}
\usepackage{lineno}

\def \nustar {{\em NuSTAR}}

\def \swift {{\em Swift}}

\def \maxi {{MAXI}}

\def \object {{MAXI J1631--479}}

\newcommand{\Msun}      {\mbox{$M_{\mathord\odot}$}}

\shortauthors{Xu et al.}

\begin{document}

\title{Studying the reflection spectra of the new black hole X-ray binary candidate MAXI J1631--479 observed by NuSTAR: A Variable Broad Iron Line Profile}

\author{Yanjun Xu\altaffilmark{1}}
\author{Fiona A. Harrison\altaffilmark{1}}
\author{John A. Tomsick\altaffilmark{2}}
\author{Dominic J. Walton\altaffilmark{3}}
\author{Didier Barret\altaffilmark{4}}
\author{Javier A. Garc\'ia\altaffilmark{1,5}}
\author{Jeremy Hare\altaffilmark{2, 6, 7}}
\author{Michael L. Parker\altaffilmark{8}}

\affil{$^{1}$ Cahill Center for Astronomy and Astrophysics, California Institute of Technology, Pasadena, CA 91125, USA}
\affil{$^{2}$ Space Sciences Laboratory, 7 Gauss Way, University of California, Berkeley, CA 94720-7450, USA}
\affil{$^{3}$ Institute of Astronomy, University of Cambridge, Madingley Road, Cambridge CB3 0HA, UK}
\affil{$^{4}$ Universit\'e de Toulouse; CNRS;  Institut de Recherche en Astrophysique et Plan\'etologie; 9 Avenue du colonel Roche, BP 44346, F-31028 Toulouse cedex 4, France}
\affil{$^{5}$ Remeis Observatory \& ECAP, Universit\"at Erlangen-N\"urnberg, Sternwartstr.~7, 96049 Bamberg, Germany}
\affil{$^{6}$ NASA Goddard Space Flight Center, Greenbelt, MD 20771, USA}
\affil{$^{7}$NASA Postdoctoral Program Fellow}
\affil{$^{8}$ European Space Agency (ESA), European Space Astronomy Centre (ESAC), E-28691 Villanueva de la Ca\~{n}ada, Madrid, Spain}

\begin{abstract}
We present results from the {\em Nuclear Spectroscopic Telescope Array} (\nustar) observations of the new black hole X-ray binary candidate \object\ at two epochs during its 2018--2019 outburst, which caught the source in a disk dominant state and a power-law dominant state. Strong relativistic disk reflection features are clearly detected, displaying significant variations in the shape and strength of the broad iron emission line between the two states. Spectral modeling of the reflection spectra reveals that the inner radius of the optically-thick accretion disk evolves from $<1.9$~$r_{\rm g}$ to $12\pm1$~$r_{\rm g}$ (statistical errors at 90\% confidence level) from the disk dominant to the power-law dominant state. Assuming in the former case that the inner disk radius is consistent with being at the ISCO, we estimate a black hole spin of $a^*>0.94$.  Given that the bolometric luminosity is similar in the two states, our results indicate that the disk truncation observed in \object\ in the power-law dominant state is unlikely to be driven by a global variation in the accretion rate. We propose that it may instead arise from local instabilities in the inner edge of the accretion disk at high accretion rates. In addition, we find an absorption feature in the spectra centered at $7.33\pm0.03$ keV during the disk dominant state, which is evidence for a rare case that an extremely fast disk wind ($v_{\rm out}=0.067^{+0.001}_{-0.004}~c$) is observed in a low-inclination black hole binary, with the viewing angle of $29\pm1^{\circ}$ as determined by the reflection modeling.
\end{abstract}

\maketitle

\section{INTRODUCTION}
The majority of Galactic black hole X-ray binaries are found as X-ray transients \citep{corr16, teta16}. These are mostly low-mass X-ray binaries (LMXBs), which are stellar-mass black holes accreting from low-mass donor stars that go into recurrent outbursts due to thermal-viscous instabilities in the accretion disc \citep[e.g.,][]{frank02}. During a typical outburst, lasting from months to years, a black hole X-ray binary  displays characteristic evolution in its X-ray spectral and timing properties, which are classified into different spectral states \citep[see][for reviews]{bhb_rev06,bell16}. The outburst usually starts from a low/hard state, transitions to a high/soft state then returns to the low/hard state at the end of the outburst. In addition, intermediate states are often found close to the time of the state transitions.

The X-ray emission from Galactic black hole X-ray binaries comes primarily from two components: blackbody radiation from the accretion disk and inverse Compton emission from the hot and tenuous corona, and they are believed to be coupled in the framework of the disk-corona model \citep[][]{haardt91,haardt93}.
Reprocessing of the hard X-ray continuum emission by the optically thick accretion disk also imprints characteristic features in the X-ray spectrum, most prominently, the ionized Fe K$\alpha$ fluorescence emission lines, the Fe K absorption edge and the Compton reflection hump arising from absorption and Compton back-scattering \citep[][]{gil88,lightman88, fabian89}. Relativistic effects at a few gravitational radii from the central black hole distort and blend the reflection features, giving rise to a broad and asymmetric line profile.

Modeling the relativistic disk reflection features helps to probe the physical conditions of inner accretion flow around the central black hole (see \citealt{miller_rev, fabian10} for reviews and references within). Notably, via fitting the spectrum with physically self-consistent reflection models, it offers a method to locate the innermost edge of the optically-thick accretion disk by measuring the degree of gravitational redshift that modifies the reflection spectrum. Assuming that the innermost disk radius is associated with the innermost stable circular orbit (ISCO) around the black hole, we can directly estimate the spin of the black hole. Recently, with high sensitivity and broadband spectral coverage, \nustar\ detected strong relativistic reflection features in several known or new black hole X-ray binaries or binary candidates, where the broad iron line profiles are resolved \citep[e.g.,][]{tomsick14,miller15, parker16, walton_v404, xu_j1535, xu_j1658, buisson19}. These high-quality datasets of bright black hole X-ray binaries are free from pile-up distortions and enable detailed studies of the inner accretion flow of black holes based on the diagnostics of relativistic disk reflection features. 

\begin{figure*}
\centering
\includegraphics[width=0.98\textwidth]{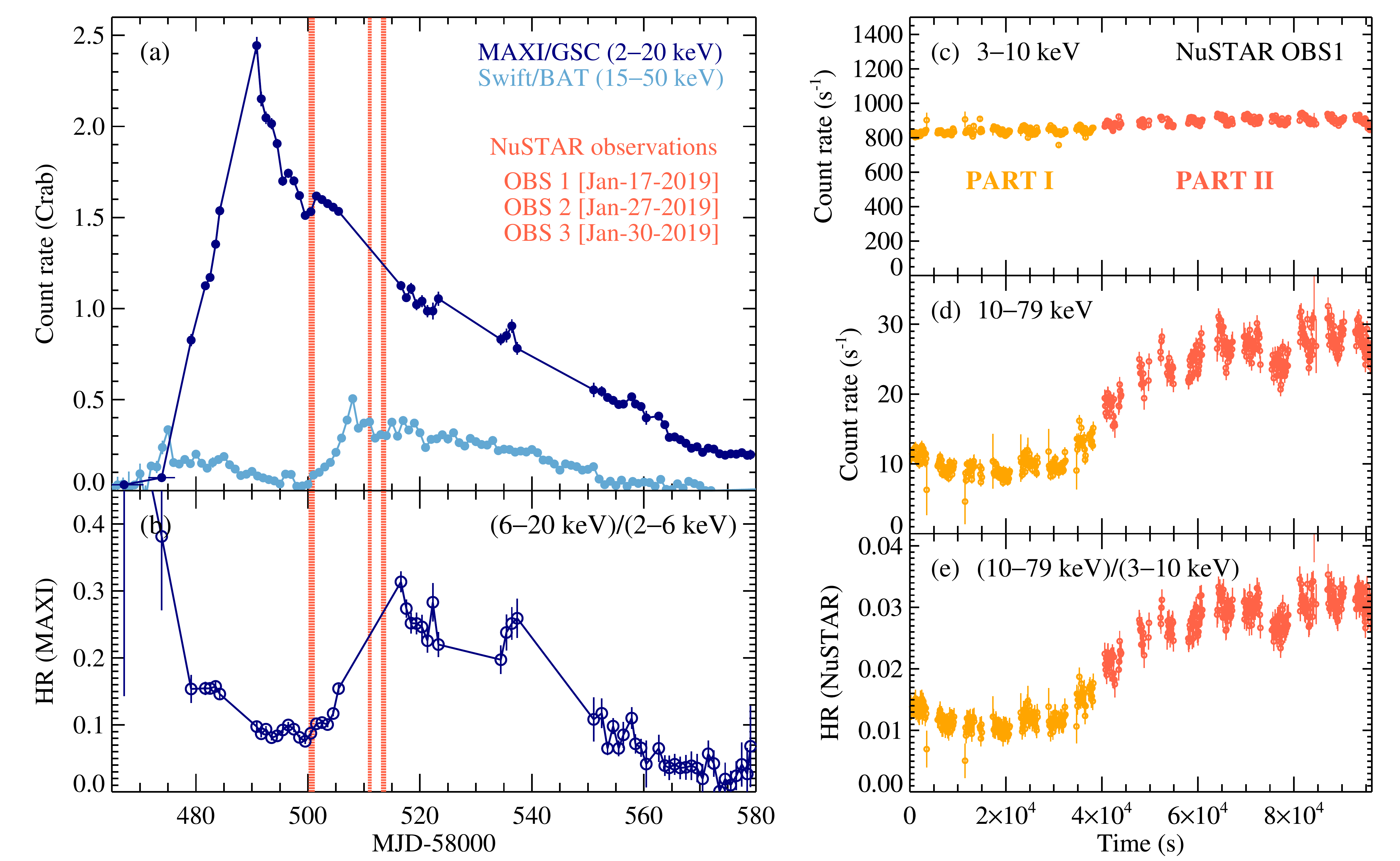}
\caption{Left panels: (a) Long-term MAXI and \swift/BAT monitoring light curves of the 2018--2019 outburst of the new black hole binary candidate \object. The flux in Crab unit is estimated from count rates in the corresponding energy bands. The orange shaded areas mark our \nustar\ observations. (b) Hardness ratio (HR) estimated by count rates in the energy bands of 2--10 keV and 10--20~keV of \maxi. Right panels: \nustar\ light curves and HR of \object\ for one module (FPMB) during OBS1, when significant flux variation is detected in the hard X-ray band. The \nustar\ count rates have been corrected for deadtime.
\label{fig:fig1}}
\end{figure*}

\object\ is a new Galactic black hole X-ray binary candidate discovered when the source went into outburst in late 2018. It is only about 8.9$\arcmin$ away from the known X-ray pulsar AX J1631.9--4752. X-ray monitoring observations by \maxi\ revealed enhanced X-ray emission from an area consistent with the position of AX J1631.9--4752 from December 2018, but greatly exceeding the maximum historically recorded flux of AX J1631.9--4752 \citep{koba18}. We performed two short \nustar\ observations on December 28, 2018 to explore the possibility of another source residing within the \maxi\ error circle (radius of 0.17$^\circ$). Indeed, \nustar\ clearly resolved two point sources within the \maxi\ error region, and confirmed that the enhanced X-ray emission came from a bright uncatalogued X-ray transient at a flux of about 0.8 Crab (2.0--10.0 keV), which was named \object\ \citep{miya18}. The characteristics of the light curve and the shape of the energy spectrum of \object\ are both typical for Galactic X-ray binaries in outburst. In addition, the lack of pulsations and the detection of a strong broad Fe K$\alpha$ emission line from the preliminary analysis of the \nustar\ data made the source a strong black hole candidate. Subsequent radio, optical and X-ray observations have been performed to further investigate the characteristics of this new black hole binary candidate \citep[e.g.,][]{russ_1631,eij_1631,kong_1631}. A radio observation on January 13, 2019 by ATCA suggests the detection of an optically-thin radio flare in a soft state black hole binary \citep{russ_1631}.

\section{Observations and DATA REDUCTION}

\begin{deluxetable}{cccc}
\tablewidth{\columnwidth}
\tablecolumns{4}
\tabletypesize{\scriptsize}
\tablecaption{\nustar\ Observations of \object\ \label{tab:tab1}}
\tablehead{
& \colhead{ObsID}
& \colhead{Start Time (UTC)}
& \colhead{Exposure (ks)}
} 
\startdata
\multirow{2}{*}{OBS1}  &\multirow{2}{*}{90501301001} &\multirow{2}{*}{Jan-17-2019  02:16} &7.5 (PART I)   \\
\noalign{\smallskip} 
& & &8.8 (PART II)   \\
\noalign{\smallskip}
OBS2  &80401316002 &Jan-27-2019 16:56 &10.1 \\
\noalign{\smallskip}  
OBS3  &80401316004 &Jan-30-2019 01:26  &14.4 
\enddata 
\tablecomments{
Exposure time is deadtime corrected on-source live time for one \nustar\ module, FPMA. Mode 6 data acounts for about 10\% of the exposure time.
}
\end{deluxetable}
\begin{figure*}
\centering
\includegraphics[width=0.98\textwidth]{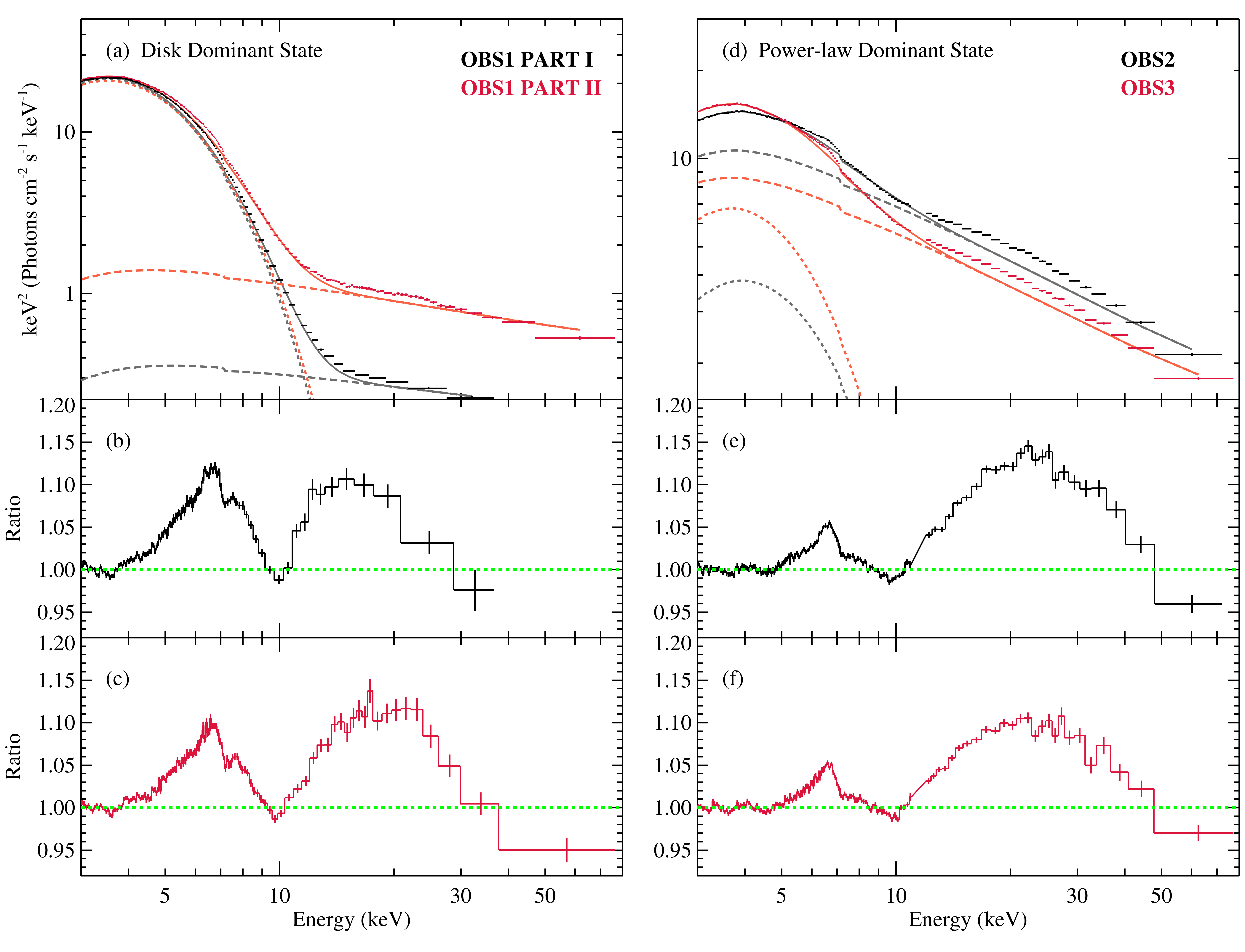}
\caption{Top panels: Unfolded \nustar\ spectra of \object\ in the disk dominant and the power-law dominant state, with the model of a disk blackbody and a non-thermal power-law component modified by neutral absorption, {\tt TBabs*(diskbb+powerlaw)} (Model 1). The disk blackbody and the power-law model component are plotted in dashed and long-dashed lines, respectively. Middle and bottom panels: Strong relativistic reflection features (i.e., a broad iron line and the Compton reflection hump) are shown in the residuals. The spectra are rebinned for display clarity.
\label{fig:fig2}}
\end{figure*}

\begin{figure*}
\centering
\includegraphics[width=0.98\textwidth]{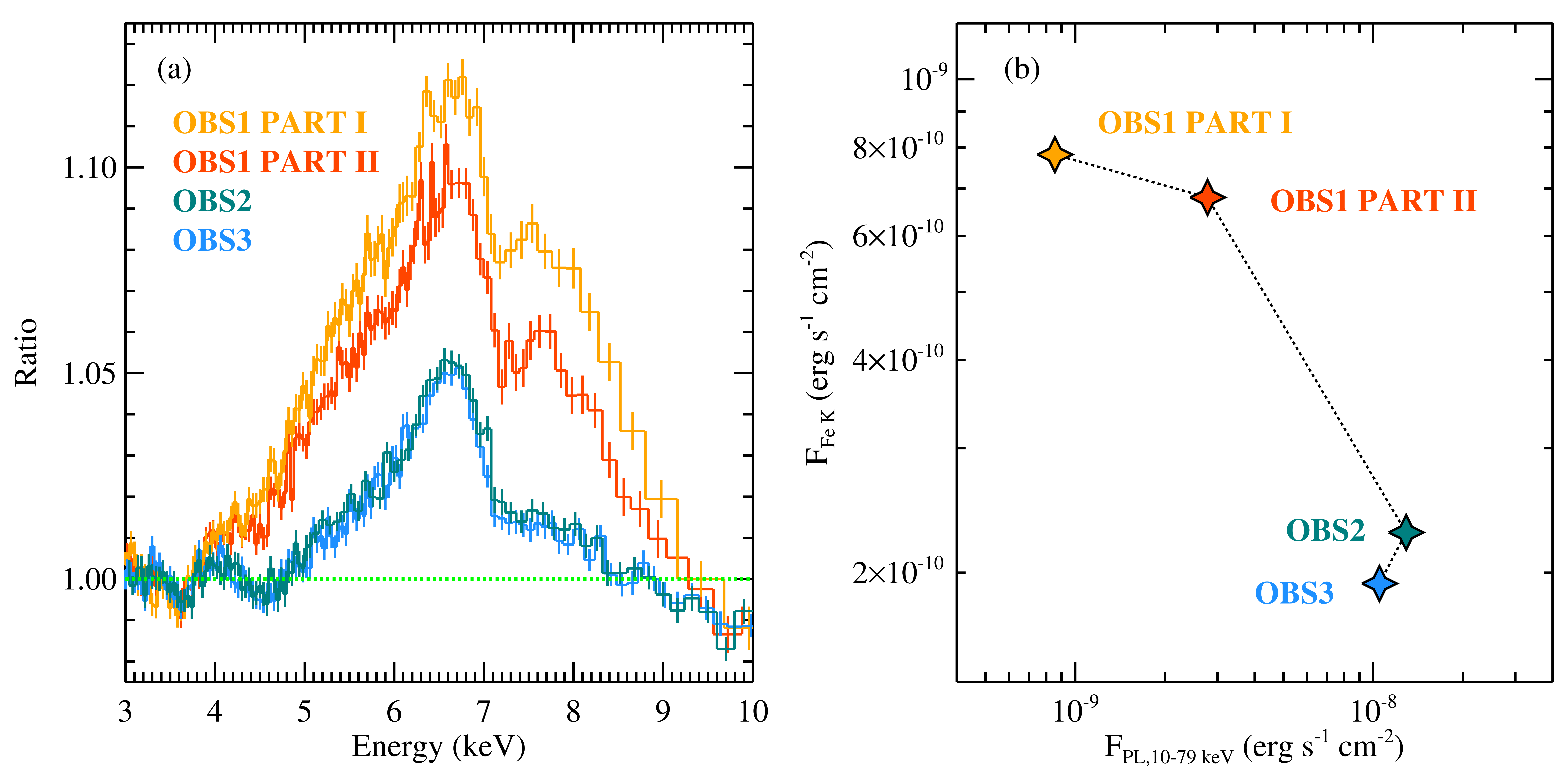}
\caption{Left panel: Zoom-in plot of the Fe K band in the spectral residuals of Figure~\ref{fig:fig2}. The spectra are rebinned for display clarity. Right panel: Evolution of the Fe K line flux with the hard X-ray flux from the corona, the later is represented by the flux of the power-law model component in the 10--79 keV band.
\label{fig:fig3}}
\end{figure*}

The new black hole binary candidate \object\ was observed by \nustar\ \citep{nustar} on January 17, January 27, and January 30, 2019 (see Table~\ref{tab:tab1} for details). We reduced the \nustar\ data following standard procedure using NuSTARDAS pipeline v.1.6.0 and CALDB v20170817. The source spectra were extracted from a circular region with the radius of 200\arcsec\ from the two \nustar\ focal plane modules (FPMA and FPMB). Corresponding background spectra were extracted using polygonal regions from source-free areas in the detectors. We also extracted spectra from mode 6 data following the procedures described in \cite{walton_cygx1} to maximize the available exposure time. Mode 6 data are taken when an aspect solution is not available from the star tracker on the optics bench (CHU4), but solutions are available from the star trackers on the main spacecraft bus (CHU1,2,3), sufficient for spectral analysis of point sources (see \citealt{walton_cygx1} for further details). \object\ was very bright during the three \nustar\ observations, the deadtime corrected count rates are $\sim$850--1000~ct~s$^{-1}$ for one module, exceeding the Crab count rate in the \nustar\ band. For spectral analysis, we coadded the FPMA and FPMB spectra from each observation using the {\tt addspec} tool in HEASOFT v6.19. The \nustar\ spectra were grouped to have a signal-to-noise ratio (S/N) of 20 per energy bin.

Significant spectral variability was found during the first \nustar\ observation (OBS1, see Figure~\ref{fig:fig1}, right panel), when the count rate in the hard X-ray band (>10 keV) increases by a factor of $\sim$3 and the count rate in the soft X-ray band (<10 keV) remains roughly constant. For spectral modeling, we separated spectra corresponding to the first and second half of the observation, noted as the OBS1 PART I and PART II henceforth. No strong spectral variation was detected during OBS2 and OBS3, which enables a time-averaged spectral analysis. We use the \nustar\ spectrum up to 40 keV for OBS1 PART I, as the spectrum starts to become background dominated above 40 keV due to the weakness of power-law tail during this period. For OBS2 and OBS3, we ignore spectra in the band of 11--12 keV due to the presence of a narrow dip centered at 11.5~keV, which is calibration related and not intrinsic to the source. The apparent dip is weak (EW$\approx$10~eV) and only noticeable in very bright sources with a hard energy spectrum. For other observations, we use \nustar\ spectra in the full energy band of 3--79~keV.

To address the spectral states during our \nustar\ observations in a broader context, we plot the duration of the \nustar\ observations with the long-term \maxi/GSC \citep{mat09} and \swift/BAT \citep{swiftbat} monitoring light curves (Figure~\ref{fig:fig1}, left panel). The {\maxi/GSC
} light curve and hardness ratio were produced by the \maxi/GSC on-demand web interface\footnote{http://maxi.riken.jp/mxondem/}. We extracted source counts from a circular region with a radius of 1.6$^\circ$ centered on the source position and extracted background counts from a region with a radius of 2$^{\circ}$, using auto bright-source exclusion with a minimum exclusion radius of 1$^\circ$. The \swift/BAT light curve was obtained from the \swift/BAT Hard X-ray Transient Monitor\footnote{https://swift.gsfc.nasa.gov/results/transients/}. We note that AX J1631.9--4752 was found to be faint ($\sim$2~mCrab in 2--10~keV) by the \nustar\ observation on December 28, 2018. Therefore, although \maxi\ and BAT cannot resolve \object\ and AX J1631.9--4752, the contamination from AX J1631.9--4752 is negligible. From the monitoring light curves, it is clear that \nustar\ OBS1 caught the source in a disk dominant state, when soft X-ray emission from the accretion disk dominates the spectrum. OBS2 and OBS3 were triggered during the phase when the source was undergoing significant spectral hardening, on the declining phase of the outburst.

\section{Spectral Modeling}
\label{sec:sec3}

We detect strong relativistic disk reflection features in the spectra of \object. To highlight the relativistic disk reflection features, we fit the \nustar\ spectrum with a disk blackbody model ({\tt diskbb}; \citealt{diskbb}) plus a power-law model modified by neutral absorption, {\tt TBabs*(diskbb+powerlaw)} (Model 1), in XSPEC notation, avoiding energy ranges corresponding to prominent reflection features (4--8~keV and 12--30 keV). In this work, we perform all spectrum modeling in XSPEC v12.9.0n \citep{xspec}, and use the cross-sections from \cite{crosssec} and abundances from \cite{wil00} in the {\tt TBabs} neutral absorption model. All uncertainties are reported at the 90\% confidence level. We fit the spectra with an absorbed disk blackbody model, {\tt TBabs*diskbb} in XSPEC. As shown in Figure~\ref{fig:fig2}, a broad and asymmetric iron line peaking at $\sim$6--7~keV and Compton hump at $\sim$20--30~keV are evident in the spectral residuals. 

Comparing the relative strength of the {\tt diskbb} and {\tt powerlaw} component in Figure~\ref{fig:fig2}, we note that OBS1 caught \object\ in a state when thermal emission from the accretion disk dominates, consistent with the canonical soft state of black hole binaries. OBS2 and OBS3 were taken when the steep power-law component is dominant, which is similar to the characteristics of the intermediate state/very high state/steep power-law state in some black hole binaries \citep[e.g.,][]{bhb_rev06,bell16}. For simplicity, we henceforth refer to the spectral state during the first and second epoch as the disk dominant and the power-law dominant state, respectively, to avoid possible ambiguities in terms of the state classification.  

The broad Fe K$\alpha$ line profile clearly varies between the two states: it appears to be broader and stronger in the disk dominant state when compared with the power-law dominant state. As shown in Figure~\ref{fig:fig3}(a), the red wing of the line extends to $\sim$4~keV and $\sim$5~keV for the disk dominant and the power-law dominant states, respectively. And the line equivalent width (EW)\footnote{EW and flux of the broad iron line is estimated by adding a {\tt Gaussian} emission line model to Model 1 with other model parameters fixed at the best-fit values in Model 1.} decreased from $\sim$180--210~eV to $\sim$70~eV from the disk dominant state to the power-law dominant state. We note that by broad Fe K$\alpha$ line here, we are referring to the blurred reflection feature that comes from the iron emission lines and the Fe K absorption edge, whose contribution to the reflection feature in the Fe K band varies with the ionization state \citep{ross05}. The shape of the broad Fe K$\alpha$ line profile is relatively constant over shorter time intervals (between OBS1 PART I and PART II, and between OBS2 and OBS3), although the spectral continuum varies considerably. Therefore, for more detailed spectral modeling, we fit the spectra of the two states separately, and link the relativistic reflection parameters between OBS1 PART I and PART II, and between OBS2 and OBS3.

We plot the evolution of the iron line flux with the power-law flux in the 10--79 keV band during our \nustar\ observations in Figure~\ref{fig:fig3}(b), which represent the strength of the observed reflection feature in the Fe K band and the strength of the hard X-ray illumination from the corona, respectively. The broad iron line flux does not correlate positively with the flux of the non-thermal power-law component, which is in contradiction to the general trend found by previous studies of black hole X-ray binaries observed by {\em RXTE} \citep[e.g.,][]{park04,rossi05,reis13,steiner16}. The ratio of the iron line flux to the power-law flux in the 10--79 keV band varies by a factor of $\sim$50 during our \nustar\ observations. Although the effects of gravitational light bending would possibly produce an anti-correlation, the effect is not predicted to be this large in the lamppost coronal geometry \citep[e.g.,][]{mini04, reis13}. The broad iron line we observed in \object\ during OBS 1 PART I is unusually strong when compared with the weakness of the coronal emission, despite that reflection features are generally believed to be weak in disk dominant states. This implies that the conventional picture that the apparent reflection features solely originate from the reprocessed non-thermal coronal emission may not be true for the case of \object\ in the disk dominant state.

\subsection{Disk Dominant State}

\begin{figure*}
\centering
\includegraphics[width=0.98\textwidth]{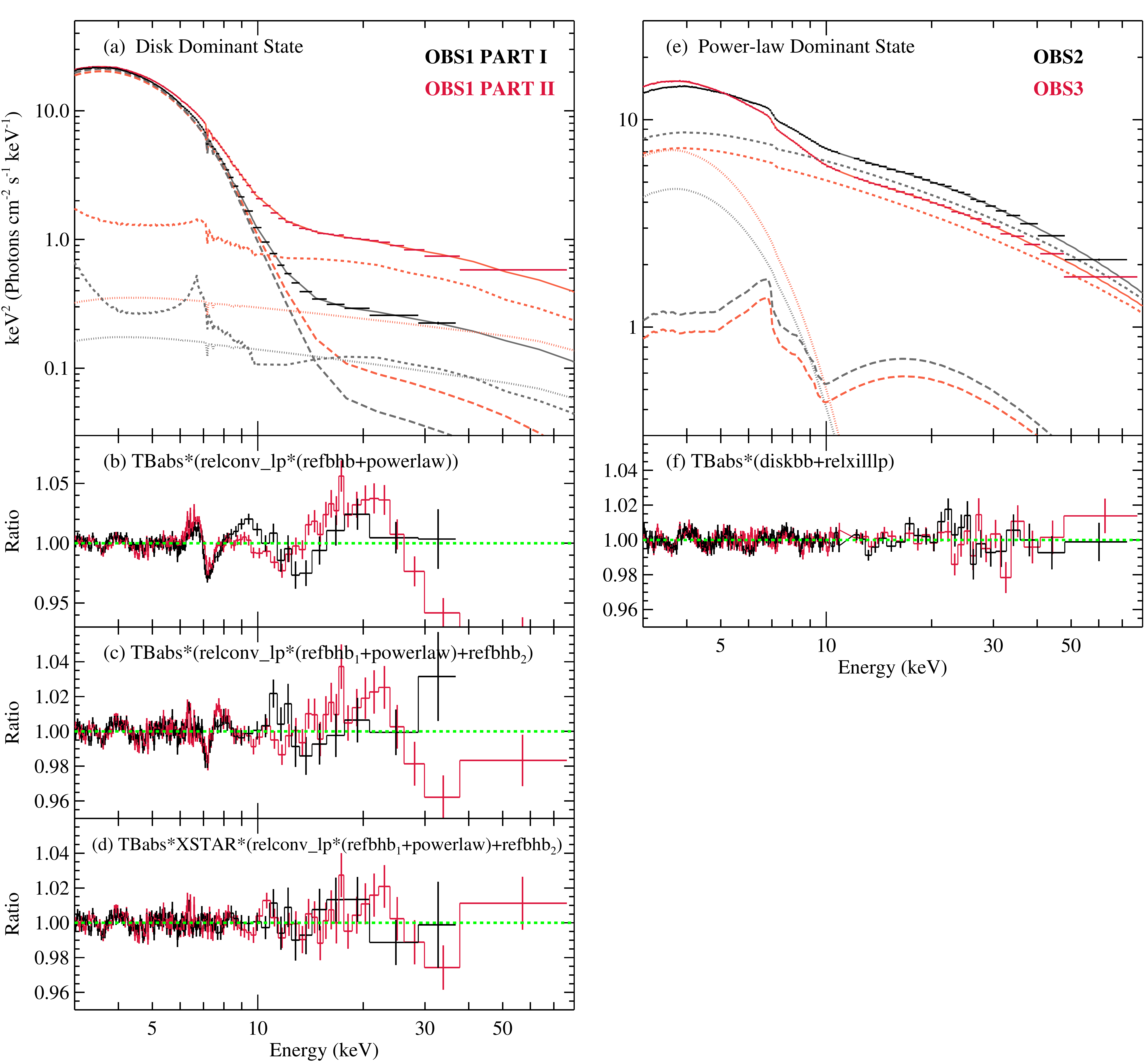}
\caption{Top panels: Unfolded \nustar\ spectra of \object\ with the best-fit models (Model 3 for the disk dominant state, Model 4 for the power-law dominant state). In panel (a), the total model is plotted in solid lines together with the non-thermal coronal emission, {\tt powerlaw} (dotted lines), the distant reflection component, {\tt refbhb$_{2}$} (dashed lines), and the thermal disk emission with relativistic disk reflection, {\tt refbhb$_{1}$} (long-dashed lines). In panel (e), the individual model components are the thermal disk emission (dotted lines), the non-thermal coronal emission (dashed lines), and the relativistic reflection component (long-dashed lines).  Bottom panels: Data/model residuals. The spectra are rebinned for display clarity.
\label{fig:fig4}}
\end{figure*}

The shape and the strength (EW) of the Fe K$\alpha$ line in \object\ remains relatively constant between OBS1 PART I and PART II, despite the large flux variation of the non-thermal emission in the hard X-ray band, implying that the strong and constant thermal disk emission plays a major role in shaping the reflected emission. In this case, the thermal disk component is dominant up to $\sim$10~keV (see Figure~\ref{fig:fig2}, left panel), which should provide the majority of the high energy X-ray photons that are able to ionize Fe K-shell electrons and produce the Fe K$\alpha$ emission line (i.e., photons with energies above 7.1~keV), when transmitting through the disk atmosphere. Thus, thermal disk photons can also cause iron fluorescence and cause an increase in the observed iron line flux above what would be expected from the coronal illumination alone. Therefore, we propose that this reprocessing of high energy disk photons could explain the anti-correlation observed in Figure~\ref{fig:fig3}(b). This would explain why the iron line flux in the disk dominant state is not correlated with the power-law flux without the need to invoke light bending effects on the coronal emission. In addition, strong disk emission should have a significant effect on determining the ionization structure of the disk atmosphere.

Therefore, in the disk dominant state of \object, thermal radiation from the hot accretion disk is important to the reflection process. We use the {\tt refbhb} \citep[][]{refbhb} reflection model to fit the reflection spectra observed here. The {\tt refbhb} reflection model takes into account the interaction between thermal disk photons and the disk atmosphere by directly including disk emission entering the surface layer from below. Thus the model is suitable for the disk dominant states of black hole binaries, when the thermal disk emission is hot and strong. The model self-consistently calculates the emergent spectrum that results from illumination of the hot inner portion of a constant density disc atmosphere and takes into account line broadening due to Compton scattering. The parameters of the model are the number density of hydrogen in the illuminated surface layer, $H_{\rm den}$, the blackbody temperature of the thermal disk emission entering the disk surface layer from below, $kT_{\rm BB}$, the power-law index of the coronal emission illuminating the surface layer from above, $\Gamma$, and the flux ratio of the coronal emission illuminating the disk and the thermal emission radiating from the disk, Illum/BB. Elemental abundances are fixed at solar values in the {\tt refbhb} model. 

\begin{deluxetable}{cccc}
\tablewidth{\columnwidth}
\tablecolumns{5}
\tabletypesize{\scriptsize}
\tablecaption{Spectral Fitting of \object\ (Disk Dominant State) \label{tab:tab2}}
\tablehead{
\noalign{\smallskip}
\multicolumn{3}{c}{Model 3:~~{\em TBabs*{\scriptsize XSTAR}*(relconv\_lp*(refbhb$_{1}$+powerlaw)+refbhb$_{2}$)}} \\
\noalign{\smallskip} 
\hline
\noalign{\smallskip} 
\colhead{Parameter}
& \colhead{OBS1(PART I)}
& \colhead{OBS1(PART II)}
} 
\startdata
 $N_{\rm H, TBabs}$ ($\rm \times10^{22}~cm^{-2}$)   &\multicolumn{2}{c}{$3.3^{+0.4}_{-0.3}$}             \\
\noalign{\smallskip}                                    
 $N_{\rm H, {\scriptsize XSTAR}}$ ($\rm \times10^{22}~cm^{-2}$)   &\multicolumn{2}{c}{$1.48^{+0.15}_{-0.12}$}       \\
\noalign{\smallskip}  
log~$({\xi})_{, \rm XSTAR}$ (log [$\rm erg~cm~s^{-1}$])        &\multicolumn{2}{c}{$4.81^{+0.18}_{-0.12}$}       \\
\noalign{\smallskip}
 $v_{\rm out, XSTAR}$ ($c$)  &\multicolumn{2}{c}{$0.067^{+0.001}_{-0.004}$}        \\
\noalign{\smallskip}
 $h$ ($r_{\rm g}$)$^{\rm a}$         &\multicolumn{2}{c}{$<3.7$}                        \\
\noalign{\smallskip}    
$a^*$ ($c\rm{J/GM^2}$)       &\multicolumn{2}{c}{$>0.94 $}                        \\
\noalign{\smallskip}
 $R_{\rm in}$ ($r_{\rm g}$)            &\multicolumn{2}{c}{$R_{\rm ISCO}^f$ ($<1.9$~$r_{\rm g}$)}                                   \\                                   
\noalign{\smallskip}
$i$ ($^\circ$)             &\multicolumn{2}{c}{$29\pm3$}                              \\
\noalign{\smallskip} 
 $\Gamma$         &$2.38^{+0.09}_{-0.04}$     &$2.32^{+0.06}_{-0.05}$                            \\
 \noalign{\smallskip}
 Norm (powerlaw)            &$9^{+12}_{-5}$ &$17^{+25}_{-11}$                                     \\
\noalign{\smallskip}               
$kT_{\rm BB, 1}$ (keV)   &\multicolumn{2}{c}{$0.94\pm0.01$}       \\

\noalign{\smallskip}                                                
 $H_{\rm den, 1}$ (cm$^{-3}$) &\multicolumn{2}{c}{$1.7^{+0.3}_{-0.4}$$\times10^{21}$}                      \\
\noalign{\smallskip}
 Illum/BB$_{\rm 1}$               &$0.04^{+0.04}_{-0.01}$    &$0.07^{+0.05}_{-0.01}$                       \\
 \noalign{\smallskip}
 Norm (refbhb$_1$)$^{\rm b}$            &$7^{+2}_{-1}$                &$10\pm1$                     \\
\noalign{\smallskip}
$kT_{\rm BB, 2}$ (keV)   &\multicolumn{2}{c}{$0.22^{+0.06}_{-0.02}$}        \\
\noalign{\smallskip}                                                           
 $H_{\rm den, 2}$ (cm$^{-3}$)        &\multicolumn{2}{c}{$4^{+13}_{-1}$$\times10^{17}$}                         \\
\noalign{\smallskip}
 Illum/BB$_{\rm 2}$            &$0.08^{+0.06}_{-0.03}$                &$0.4^{+0.3}_{-0.1}$                     \\
\noalign{\smallskip}
 Norm (refbhb$_2$)            &$0.7\pm0.1$                &$2.6^{+0.2}_{-0.3}$                     \\

\noalign{\smallskip}
\hline                                                                                                        
\noalign{\smallskip}                                            $\chi^2/{\nu}$                        &\multicolumn{2}{c}{1027.9/791=1.30}     \\
\noalign{\smallskip}
\hline
\noalign{\smallskip}
$F_{\rm 3-10~keV}$~(erg~cm$^{-2}$~s$^{-1}$)$^{\rm c}$  &$2.5\times10^{-8}$  &$2.6\times10^{-8}$  \\
$F_{\rm 10-79~keV}$~(erg~cm$^{-2}$~s$^{-1}$)$^{\rm c}$ &$9.7\times10^{-10}$ &$2.9\times10^{-9}$  \\
$F_{\rm 0.1-100~keV}$~(erg~cm$^{-2}$~s$^{-1}$)$^{\rm d}$  &$1.7\times10^{-7}$ &$1.6\times10^{-7}$ 
\enddata  
\tablecomments{
Parameters marked with a superscript $f$ are fixed during the spectral fitting. $^{\rm a}$ $r_{\rm g}\equiv{\rm GM}/c^2$, is the gravitational radius. $^{\rm b}$ The {\tt refbhb} model is normalized based on the power-law incident flux in 1--100 keV. $^{\rm c}$ Observed flux in the corresponding energy bands. $^{\rm d}$ Flux corrected for absorption. 
}                                                                                
\end{deluxetable}

We convolve the disk reflection component with the {\tt relconv\_lp} model \citep{dauser10,dauser13} to measure the relativistic blurring effects on the reflection component. The {\tt relconv\_lp} model assumes an idealized lamp-post geometry (i.e., the corona is a point source located on the spin axis of the black hole above the accretion disk). It parameterizes the disk emissivity profile by the height of the corona, $h$, with a lower $h$ corresponing to a steeper disk emissivity profile \citep[e.g.,][]{wilkins12,dauser13}. We note that even in cases where the lamp-post geometry may not be the most realistic assumption, $h$ can still be viewed as a proxy for the general shape of disk emissivity profile \citep[][]{dauser13}, as currently even very high quality data cannot distinguish between the disk emissivity calculated in the lamp-post geometry and that assumed in the form of a broken power-law \citep[e.g.,][]{miller15,xu_j1535, parker15}. Other parameters of the {\tt relconv\_lp} model are the black hole spin, $a^*$, the inner radius, $R_{\rm in}$, and the inclination, $i$, of the accretion disk. 

The total model is set up in XSPEC as {\tt TBabs*} {\tt (relconv\_lp*(refbhb+powerlaw))} (Model 2). The {\tt powerlaw} model component is used here to account for the coronal emission that goes directly towards the observer without being reflected. We link the power-law index, $\Gamma$, in {\tt powerlaw} and {\tt refbhb}. As the black hole spin, $a^*$, and inner disk radius, $R_{\rm in}$, are degenerate parameters, we fix $R_{\rm in}$ at the radius of the ISCO, aiming to obtain a measurement of the black hole spin. All parameters are linked between OBS1 PART I and PART II, except $\Gamma$, Illum/BB, and the normalization of the model components.

As shown in Figure~\ref{fig:fig4}(b), Model 2 is able to account for most of the reflection features and greatly improves the fit ($\chi^2/\nu=1825.8/800$, where $\nu$ is the number of degrees of freedom), leaving only a narrow Fe K$\alpha$ emission line centered at $6.50\pm0.03$~keV, some excess in the Compton hump region, and a narrow absorption feature centered at $7.33\pm0.03$~keV in the spectral residuals. The narrow absorption feature can also be clearly seen in Figure~\ref{fig:fig3}(a), as a dip in 7--8 keV superposed on the broad iron line profile. We add an unblurred reflection component to fit the weak and narrow iron emission line, which possibly originates from distant reprocessing of the hard X-ray photons.  Absorption features in the Fe K band are commonly associated with blueshifted Fe {\small XXV}/Fe {\small XXVI} lines, but the narrow absorption line complex cannot be resolved by \nustar. They are believed to arise from absorption by outflowing material launched from the accretion disk \citep[e.g.,][]{ponti12,diaz16}. If the absorption feature at $7.33\pm0.03$~keV is associated with blueshifted He-like Fe {\small XXV} (6.70 keV), it requires an outflowing velocity of $0.094\pm0.004~c$; or a lower velocity of $0.052\pm0.004~c$ if identified with the more ionized H-like Fe {\small XXVI}. For a physical modeling of the absorption feature, we include an ionized absorption table model calculated by the {\tt XSTAR} photoionization code \citep{xstar}. We use the same {\tt XSTAR} grid as constructed for the soft state of Cygnus X-1 used in \cite{tomsick14}, the free model parameters are the absorption column density, $N_{\rm H, XSTAR}$, the ionization parameter, $\xi$, and the outflowing velocity, $v_{\rm out}$. We set up the model in XSPEC as {\tt TBabs*{XSTAR}*}{\tt (relconv\_lp*(refbhb$_1$}{\tt+powerlaw)} {\tt +refbhb$_2$)} (Model 3). 

\begin{figure}
\centering
\includegraphics[width=0.46\textwidth]{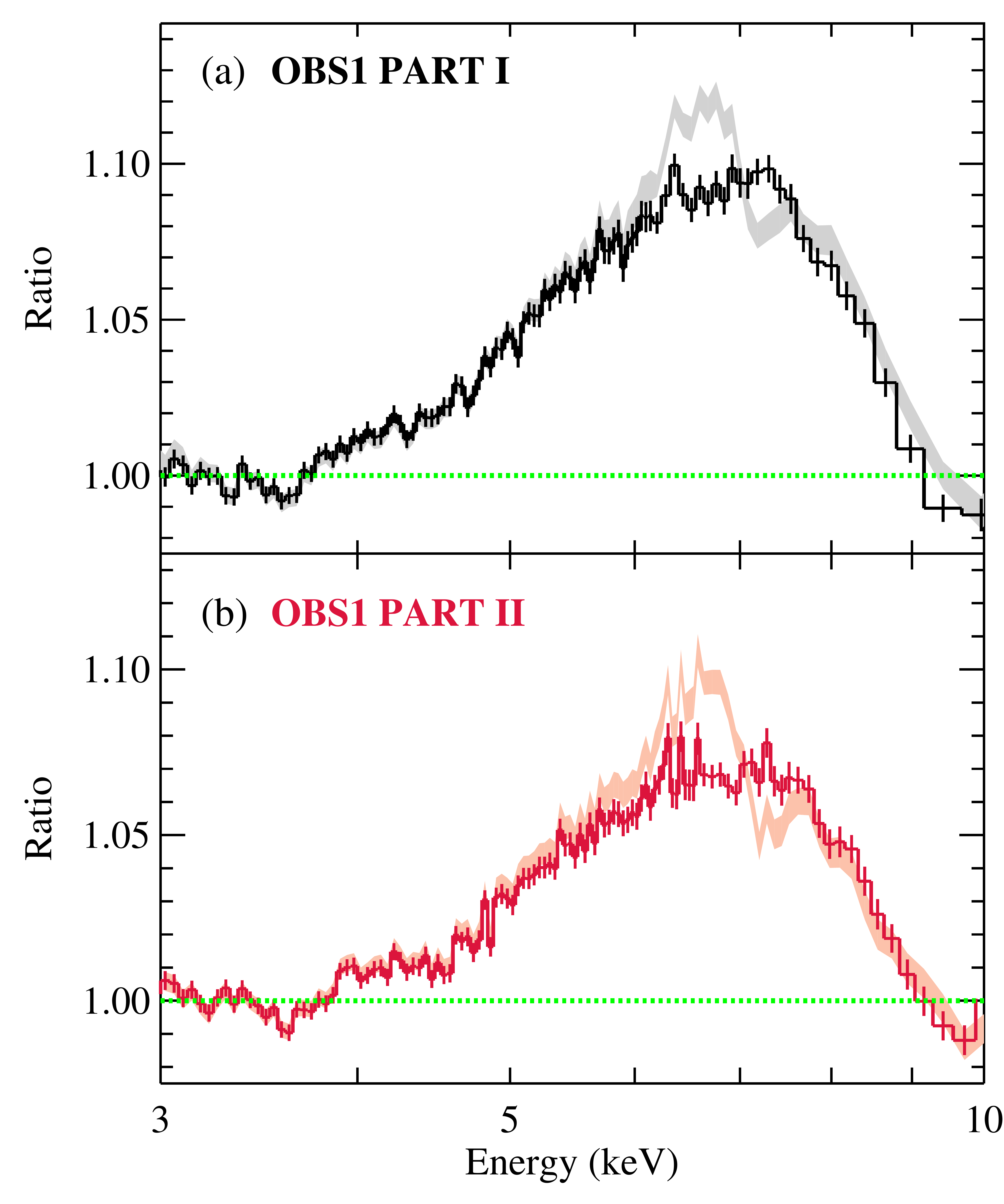}
\caption{The broad iron line profiles of \object\ observed during the disk dominant state, corrected for the absorption feature at $\sim$7.33~keV and the narrow iron emission line at $\sim$6.50~keV. The spectral residuals in the Fe K band here are estimated by adding the {\tt XSTAR} and {\tt refbhb$_2$} components to Model 1, with their parameters fixed at the best-fit values in Table \ref{tab:tab2}, and fitting the \nustar\ spectra in 3--79~keV avoiding the energy range of 4--9~keV. For comparison, the line profiles without this correction are plotted in shaded regions.
\label{fig:fig5}}
\end{figure}

Model 3 adequately describes the data, leaving no obvious structures in the residuals ($\chi^2/\nu=1027.9/791$, see Figure~\ref{fig:fig4}(d)). The addition of an {\rm XSTAR} grid improves the fit by $\Delta\chi^2\sim160$, indicating that the narrow absorption feature is significantly detected (see the comparison between Figure~\ref{fig:fig4}(c) and (d)). The best fit reveals an outflowing velocity of $v_{\rm out, XSTAR}=0.067^{+0.001}_{-0.004}~c$ for the ionized absorber, which is extremely fast for a disk wind detected in a black hole X-ray binary \citep[e.g.,][]{king12, miller15}. Mostly driven by the clear broad iron line profile, the black hole spin and physical parameters related to the inner accretion flow are well constrained (see Table~\ref{tab:tab2}). The fitting results reveal that the central compact object in \object\ is a rapidly spinning black hole with the spin parameter of $a^*>0.94$, and the inner accretion disk is viewed at a low inclination of $i=29^{\circ}\pm3^{\circ}$. By fixing the spin parameter at the maximum value of 0.998 and instead fitting for the inner disk radius, we can get an estimate of the inner disk radius of $R_{\rm in}<1.9$~$r_{\rm g}$ ($r_{\rm g}\equiv{\rm GM}/c^2$, is the gravitational radius). The
ionization state of the gas is self-consistently derived by radiative transfer equations in the {\tt refbhb} model, which is determined by the disk temperature, $kT_{\rm BB}$, the disk density, $H_{\rm den}$, and strength of the illuminating flux from the corona, Illum/BB. We can also get a reasonable constraint on the disk density and temperature based on the spectral modeling. The values are high for the relativistic reflection component, {\tt refbhb$_1$}, and decrease significantly for the unblurred reflection component,  {\tt refbhb$_2$} (see best-fit parameters in Table~\ref{tab:tab2}). This is in accordance with the unblurred reflection occurring at a large distance from the black hole, as both the disk density and temperature are predicted to drop with increasing disk radius in the standard \cite{shakura73} disk model.

For the spectral fitting above, we linked the disk temperature and the parameters about the relativistic blurring effects between OBS1 PART I and PART II, which is motivated by the constant thermal disk component and the similarity of the observed broad iron line profile. Allowing these parameters to have different values between epochs does not cause any significant change to the fitting results, therefore we keep them linked to obtain tighter constraints on the key physical parameters of interest. In this case, the strong and hot thermal disk component has a major role in shaping the reflection features rather than non-thermal emission from the corona, so the relative strength of the different non-thermal components above 20 keV cannot be well constrained (e.g., see the poorly constrained normalization of the {\tt powerlaw} model in Table~\ref{tab:tab2}). Therefore, we do not discuss the physical implications about the fraction of the power-law emission that is reflected by the inner disk, the distant reprocessing material, and the emission that goes directly to the observer as inferred from the best-fit parameters, to avoid over-interpreting the data.

We note that within 3$\sigma$ errors, the flux of the distant reflection component can be either higher or lower than that of the direct power-law component above 20 keV, but  it is always constrained to be higher than that of the relativistic reflection component. The changes to the quality of the fits and the fit parameters are only negligible if we force the power-law flux to be higher than the flux of {\tt refbhb$_2$} above 20 keV. We have tried replacing the {\tt powerlaw} component in Model 3 with a power-law model with a high-energy cutoff, {\tt cutoffpl}, a power-law model with a low-energy cutoff, {\tt expabs*powerlaw}, and a Comptionization model with both low-energy and high-energy rollovers, {\tt nthcomp} \citep{zdz96, zyc99}, which do not cause any significant difference to the fitting results and the cutoff is not required for the spectral continuum. It is difficult to understand the geometry that could cause the distant reflection to receive more coronal illumination than the relativistic one. We are currently uncertain if this is due to oversimplications in the current version of the {\tt refbhb} model, e.g., the fixed iron abundance and the self-consistently calculated ionization assuming a constant accretion rate and disk surface radius \citep{refbhb}, which might cause a bias in the relative strength of the iron line and Compton reflection hump and make the origin of the {\tt refbhb$_2$} component questionable. If this is the case, the role of the {\tt refbhb$_2$} model component in the spectral fitting might only be phenomenological in some sense. However, as shown in Figure~\ref{fig:fig5}, {\tt refbhb$_2$} only contributes to the narrow core of the iron line profile, thus would not affect the base of the broad iron line profile that we use to deduce the black hole spin.

The single-temperature blackbody description of the accretion disk in the {\tt refbhb} model is not as realistic as the multi-color blackbody disk model. As discussed in \cite{reis08}, this simplification might lead to biased values for the inner disk temperature, but would not have a significant effect on the key physical parameters determined by the reflection modeling, such as the inner radius  of the accretion disk (or the black hole spin). Modeling the \nustar\ spectra of \object\ above 4 keV, mostly only covering the Wien tail of the blackbody distribution, leads to similar fitting results, therefore we stress that the measurement of a rapid black hole spin here is robust.

For comparison, we also made an attempt to fit the data with a disk reflection model that does not directly include thermal disk emission, {\tt relxilllp} (see detailed description of the model in Section 3.2). We add a {\tt diskbb} model to account for the thermal disk component. The fit leads to extreme parameters, a high iron abundance, $A_{\rm Fe}\sim5$, and a reflection fraction, $R_{\rm ref}>10$, for OBS1 PART I, which is driven by the strong iron line but such high reflection fraction is probably unphysical. And the model fails to describe the data, leaving prominent spectral residuals.

\subsection{Power-law Dominant State}

\begin{deluxetable}{ccc}
\tablewidth{\columnwidth}
\tablecolumns{3}
\tabletypesize{\scriptsize}
\tablecaption{Spectral Fitting of \object\ (Power-law Dominant State) \label{tab:tab3}}
\tablehead{
\multicolumn{3}{c}{Model 4:~~\em{ TBabs*(diskbb+relxilllp)}} \\
\noalign{\smallskip} 
\hline
\noalign{\smallskip} 
\colhead{Parameter} 
& \colhead{OBS2}
& \colhead{OBS3}
} 
\startdata
 $N_{\rm H, TBabs}$ ($\rm \times10^{22}~cm^{-2}$)  &\multicolumn{2}{c}{$3.84^{+0.10}_{-0.09}$}        \\
\noalign{\smallskip}                                    
 $h$ ($r_{\rm g}$)       &\multicolumn{2}{c}{$4.3^{+0.5}_{-0.3}$}       \\
\noalign{\smallskip}    
$a^*$ ($c\rm{J/GM^2}$)   &\multicolumn{2}{c}{$0.96^f$}       \\
\noalign{\smallskip}
 $R_{\rm in}$ ($r_{\rm g}$)   &\multicolumn{2}{c}{$12\pm1$}                 \\                                   
\noalign{\smallskip}
$i$ ($^\circ$)     &\multicolumn{2}{c}{$29\pm1$}                  \\
\noalign{\smallskip} 
 $\Gamma$      &$2.45\pm{0.02}$  &$2.51\pm{0.02}$                  \\
\noalign{\smallskip}               
{$E_{\rm cut}$~(keV)}   &$110\pm11$   &$156\pm22$          \\
\noalign{\smallskip}
 log~$({\xi})_{,\rm ref}$ (log [$\rm erg~cm~s^{-1}$])  &\multicolumn{2}{c}{$3.56^{+0.15}_{-0.09}$}                   \\
\noalign{\smallskip}                                                
 {$A_{\rm Fe}$ (solar)}     &\multicolumn{2}{c}{$1.0^{+0.3}_{-0.1}$}                    \\
\noalign{\smallskip}                         $R_{\rm ref}$           &\multicolumn{2}{c}{0.49}                       \\
\noalign{\smallskip}   
Norm (relxilllp)       &$0.8\pm0.1$  &$0.8\pm0.1$ \\
\noalign{\smallskip}    
$kT_{\rm in}$ (keV)       &$1.19^{+0.01}_{-0.02}$  &$1.14\pm0.01$   \\
\noalign{\smallskip}   
Norm (diskbb)       &$454^{+32}_{-20}$  &$859^{+38}_{-27}$  \\
\noalign{\smallskip}                                                
\hline                                                                                                        
\noalign{\smallskip}      $\chi^2/{\nu}$   &\multicolumn{2}{c}{1669.6/1491=1.12}   \\
\noalign{\smallskip}
\hline
\noalign{\smallskip}
$F_{\rm 3-10~keV}$~(erg~cm$^{-2}$~s$^{-1}$)  &$2.3\times10^{-8}$ &$2.3\times10^{-8}$ \\
$F_{\rm 10-79~keV}$~(erg~cm$^{-2}$~s$^{-1}$) &$1.3\times10^{-8}$ &$1.1\times10^{-8}$  \\
$F_{\rm 0.1-100~keV}$~(erg~cm$^{-2}$~s$^{-1}$)  &$2.5\times10^{-7}$ &$2.5\times10^{-7}$
\enddata  
                                                           
\end{deluxetable}

\begin{figure*}
\centering
\includegraphics[width=0.92\textwidth]{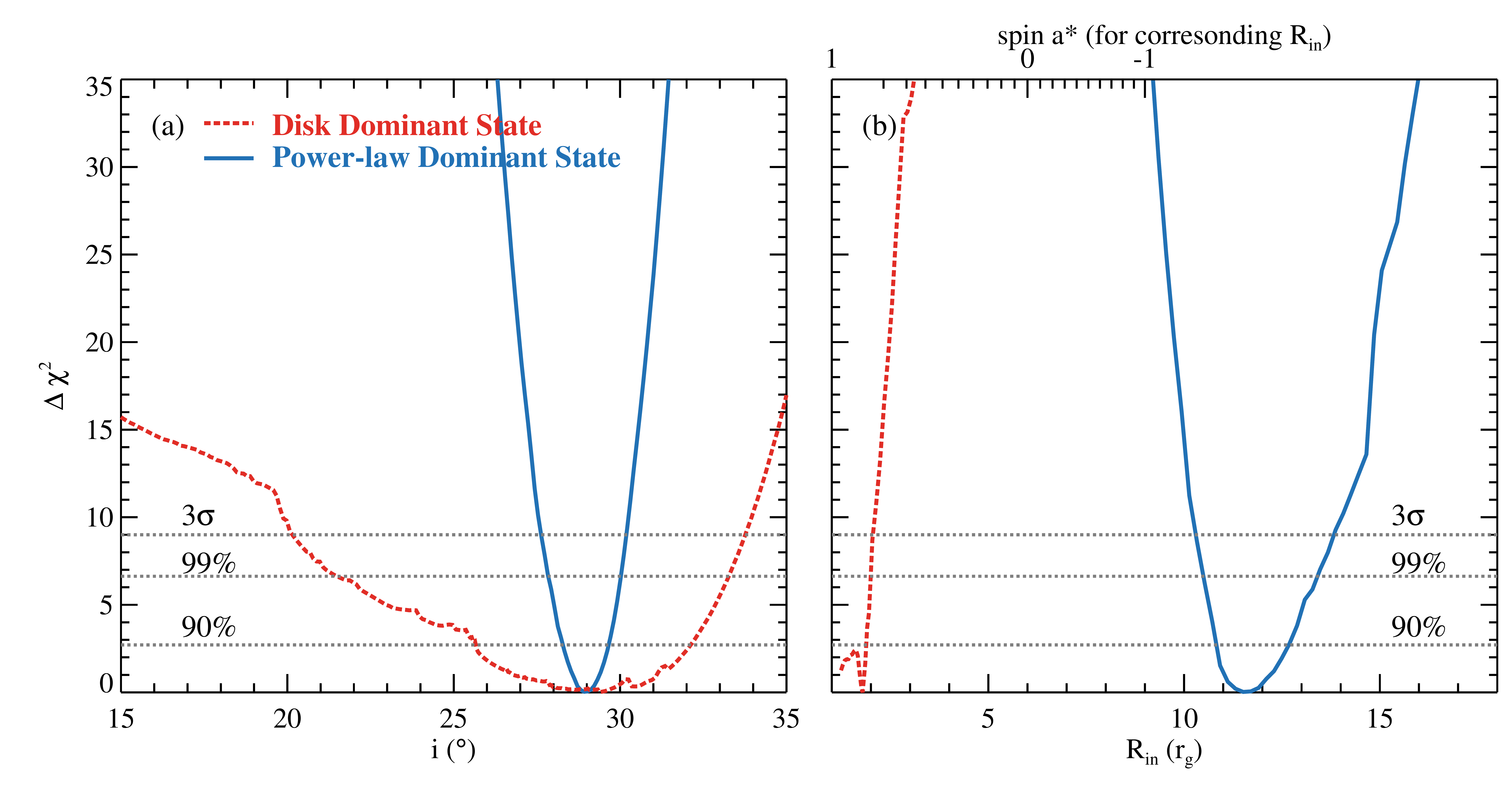}
\caption{$\Delta \chi^2$ plots for the black hole spin parameter a$^*$ and inner accretion disk radius R$_{\rm in}$. The constraints for the disk dominant state and power-law dominate state are marked in red dashed and blue solid lines. The horizontal lines indicate the 90\%, 99\%, and 3$\sigma$ confidence levels for one parameter of interest.
\label{fig:fig6}}
\end{figure*}

For the power-law dominant state, the flux of the illuminating coronal emission is significantly higher than that of the thermal disk emission, which exceeds the range of parameters covered in the {\tt refbhb} model. We instead use the {\tt relxilllp} model ({\tt relxill}; \citealt{relxilla, relxillb}) to fit the reflection spectra, which is a combination of the ionized disk reflection model {\tt xillver} \citep{garcia13}, which implements the {\tt XSTAR} code \citep{xstar} to solve the ionization structure of the disk atmosphere, and the relativistic blurring kernel {\tt relconv\_lp} \citep[][]{dauser10, dauser13}. The {\tt relxilllp} model intrinsically includes the illuminating corona emission in the shape of a power-law with an exponential high-energy cutoff, $E_{\rm cut}$. The {\tt xillver} reflection model is calculated based on the assumption of a slab geometry for the accretion disk, and does not include thermal disk photons entering the surface layer of the accretion disk from below. The model assumes a constant density ($H_{\rm den}=10^{15}$~{cm$^{-3}$}) for the surface layer of the accretion disk with the disk temperature fixed at 10 eV. Therefore, the model neglects the role of thermal disk emission on determining the physical conditions in the disk atmosphere and thus the emergence reflection spectrum, which suits the case of cool accretion disks in AGNs. The model underestimates the effect of a hot accretion disk on the reflection spectrum for black hole binaries. But it seems to be a reasonable simplification for black hole binaries in the hard and intermediate states, when the thermal disk emission is weak and non-thermal power-law component dominates the spectrum. The iron abundance, $A_{\rm Fe}$ (in solar units), and the ionization parameter $\xi$ (defined as $\xi=4\pi F_{\rm x}/n$, where $F_{\rm x}$ is the illuminating flux and $n$ is the gas density) are free parameters in the {\tt xillver} model.

In addition, we include a separate multicolor disk blackbody component, {\tt diskbb}, to fit for the thermal emission from the accretion disk. We set up the total model in XSPEC as {\tt TBabs*(diskbb+relxilllp)} (Model 4). We note that the broad iron line becomes narrower in the power-law dominant state (see Figure~\ref{fig:fig3}(a)), implying that the gravitational redshift is weaker in the line emission region, which implies that the Fe K$\alpha$ emission line is produced at a larger disk radius than that in the disk dominant state. Therefore, we fix the black hole spin at the maximum value of $a^*=0.96$ (best-fit value in the disk dominant state), and instead fit for the inner disk radius, $R_{\rm in}$. We note that the choice of the fixed spin parameter here would not affect the determination of other model parameters. If the accretion disk is truncated at a larger radius than the ISCO, the data would not be sensitive to the black hole spin parameter, as the frame-dragging effect from black holes of different spins would only cause very small differences to the emission line profile produced at large disk radii \citep[e.g.,][]{dauser13}. Model 4 fits the \nustar\ spectra well and leaves no prominent structures in the spectral residuals ($\chi^2/\nu=1669.6/1491=1.12$, see Figure~\ref{fig:fig4}(f)). The fitting results confirm that the source is not in a canonical hard state during the time of the \nustar\ observation, despite the significant spectral hardening observed in the monitoring light curves in Figure~\ref{fig:fig1}(b). From the best-fit parameters (see Table~\ref{tab:tab3}), we find the disk temperature is still high and the spectral slope remains soft in the power-law dominant state. The spectral hardening is caused by a large change in the relative strength of the thermal disk and the power-law component, the disk fraction of the total flux in 0.1--100~keV decreases from $\sim60$\% to $\sim10$\% from the disk dominant to the power-law dominant state.

In terms of the reflection parameters, the best-fit results of Model 4 (see Table~\ref{tab:tab3}) indicate significant truncation of the inner edge of the accretion disk at $R_{\rm in}=12\pm1$~$r_g$, larger than the ISCO radius for any given black hole spin (the ISCO radius depends on the black hole spin, $R_{\rm ISCO}=1.235$~$r_{\rm g}$ for $a^*=0.998$, $R_{\rm ISCO}=6$~$r_{\rm g}$ for $a^*=0$, and $R_{\rm ISCO}=$ $\sim$9~$r_{\rm g}$ for $a^*=-0.998$; \citealt{bard72, thorne74}). The disk inclination determined by the reflection method is driven by the relativistic Doppler shift of iron line, which is sensitive to the blue wing of the line profile \citep[e.g.,][]{brenn06}. The inclination of the inner accretion disk is measured to be $29\pm1^{\circ}$, which agrees well with the value obtained in the disk dominant state from the {\tt refbhb} model. The iron abundance measured by the {\tt relxilllp} model is consistent with the solar value, $A_{\rm Fe}=1.0^{+0.3}_{-0.1}$, which is also in agreement with the value of elemental abundances fixed in the {\tt refbhb} model. We note that certain systematic uncertainties could be introduced when fitting the spectra with different reflection models \citep[e.g., see the discussion in][]{middleton16}, but the well agreement we find in $i$ and $A_{\rm Fe}$ determined by the {\tt refbhb} and {\tt relxilllp} reflection models indicates such uncertainties are probably minimal here. 

In addition, we find a reflection fraction of $R_{\rm ref}=0.49$ for the power-law dominant state, significantly lower when compared with that of other black hole binaries displaying strong reflection features \citep[e.g.,][]{xu_j1535,xu_j1658,miller_j1535}. In the {\tt relxilllp} model, the value is defined as the ratio of the coronal intensity illuminating the disk to that reaching the observer, self-consistently calculated by the model in the lamp-post geometry \citep{dauser16}. Leaving the reflection fraction as a free parameter does not cause any significant change to the fit, indicating a truncated accretion disk is indeed required by the shape of broad iron profile. The low reflection fraction is consistent with the relative weakness of the broad iron line in the power-law dominant state, which is a natural outcome from the scenario where the accretion disk is truncated, as X-ray photons would be lost in the gap between the black hole and the inner edge of the truncated accretion disk, without being reprocessed by the optically-thick accretion disk.

\section{Discussion}
\label{sec:sec4}
We have performed detailed modeling of the reflection spectra of the new black hole binary candidate \object\ during its 2018--2019 outburst, which displays clear variation in the Fe K$\alpha$ emission line profile between the two epochs observed by \nustar. The inclination of the inner part of accretion disk is measured to be low in \object\ ($i\approx29^{\circ}$, closer to being face-on than edge-on). Therefore, we probably have a relatively clear view of the inner regions around the black hole without being blocked by e.g., a geometrically-thick accretion disk. Detailed modeling of the reflection spectra suggests that the variation of the broad iron line profile is caused by a change in the radius of the inner edge of the optically thick accretion disk, $R_{\rm in}$, from $<1.9$~$r_{\rm g}$ to $12\pm1$~$r_{\rm g}$, which consistently explains the changes in the iron line width and strength (see constraints of the inclination and the inner disk radius in Figure~\ref{fig:fig6}).

Assuming that the small inner disk radius measured during the first epoch is associated with the ISCO radius of the black hole, we can get an estimate of the spin of the black hole of $a^*>0.94$, close to the maximum value. The high spin measured indicates that the central compact object in \object\ is likely a black hole instead of a neutron star. In accreting neutron stars, the inner edge of accretion disk is observed to be truncated at a typically larger radius of $\sim$6--15~$r_{\rm g}$, due to either the magnetic field of the neutron star or the existence of a solid stellar surface \citep[e.g.,][]{cackett10, lu17, ludlam19}. During the second epoch, however, the inner radius measured via the reflection method becomes significantly larger, which suggests a change in the accretion mode in \object: the innermost part of the optically-thick accretion disk is replaced by an optically-thin accretion flow, which is unable to produce clear reflection features inside of the disk truncation radius. 

\subsection{The Iron Line Profile and Disk Truncation}
The inner edge of the accretion disk around black holes is predicted to vary with the mass accretion rate, which is inferred from the source luminosity, in units of the Eddington luminosity. The mass accretion rate is thought to influence the disk surface density, and thus determine the transitional inner edge of the accretion disk where the matter becomes thin to Thomson scattering, defined as the ``reflection edge'' \citep[e.g.,][]{krolik02}. When the accretion rate is high, the optically-thick accretion disk is predicted to reach down to the ISCO, and its innermost part would be replaced by an optically-thin advection-dominated accretion flow (ADAF) at low accretion rates \citep[e.g.,][]{esin97, esin98}. Different theoretical models have been proposed to discuss the mechanisms facilitating the disk truncation \citep[e.g.,][]{narayan95, honma96, meyer00, yuan04}. However, what triggers disc truncation in accreting black hole systems and the typical disk truncation radius is still debated.  

From observations, the outburst of a black hole binary is considered ideal to explore the relationship between the accretion disk inner disk radius and the accretion rate, as its X-ray luminosity varies by several orders of magnitude during a typical outburst. Recently, there have been numerous observational campaigns of black hole X-ray binaries aiming to use the reflection method to track the evolution of the inner accretion disk radius \citep[e.g.,][]{fuerst_gx339, walton_cygx1, walton_v404, xu_j17091, garcia19, kara19, buisson19}. However, in most cases, the change in the inner disk radius measured at multiple epochs is only marginally significant, and can often be considered as consistent within errors. The best cases regarding variations in the Fe K$\alpha$ line profile are the narrow lines detected in the very low states of {GX 339--4} \citep{tomsick09} and {V404 Cygni} \citep{motta17}, given that relativistically broadened lines are known to be present in their high states \citep[e.g.,][]{parker16, walton_v404}. We note that in the previous cases the Fe K$\alpha$ lines are sufficiently narrow and symmetric that they do not require general relativity effects to explain, signaling a line emission region at a large distance of $\gtrsim 10^2-10^3$~$r_{\rm g}$. As proposed in \cite{motta17}, it is possible that they arise from reprocessing by distant obscuring material, making them an analogue to obscured AGNs. Thus it is debated whether they constrain the inner edge of the accretion disk. The possibility of the existence of two reflection zones (inner accretion disk and distant reprocessing material) is also supported by the narrow core found on top of the broad Fe K$\alpha$ emission line profile in several black hole binaries  \citep[e.g.,][]{walton_v404, miller_j1535, tomsick18, xu_j1535, xu_j1658}. In general, there still lacks high S/N detections of a relativistically broadened Fe K$\alpha$ emission line changing in line width, which could be used as a strong evidence for a variable inner accretion disk radius in a black hole X-ray binary.

In this work, we obtained high S/N \nustar\ spectra of the new black hole binary candidate \object\, revealing a broad Fe K$\alpha$ emission line variable in line width and strength. Spectral modeling suggests a change in the inner disk radius from $R_{\rm in}<1.9~r_{\rm g}$ to $R_{\rm in}=12\pm1~r_{\rm g}$ between the two epochs observed by \nustar, despite the small variance in the bolometric luminosity\footnote{The bolometric luminosity is estimated from the absorption corrected flux in 0.1--100~keV from the best-fit spectral models.} of the source at $L_{\rm bol}\sim(1.3\times10^{39}-1.9\times10^{39})\times(D/8~{\rm kpc})^2$~erg~s$^{-1}$. The Eddington rate at the time of the observations is uncertain, as both the black hole mass and distance are currently unknown. Assuming a typical source distance of 8~kpc and black hole mass of 10~\Msun, we estimate the Eddington rate of $L/L_{\rm Edd}\sim(100\%-150\%)\times(10~\Msun/M)\times(D/8~{\rm kpc})^2$. Our spectral analysis of \object\ indicates that it is possible for the accretion disk to become significantly truncated at this  high accretion rate. In addition, the results suggest that in the case of \object, the change in the accretion mode between the two epochs is not driven by a variation in the global accretion rate of the system.

At high accretion rates, it is predicted that disk truncation could be triggered by thermal instabilities due to the dominance of radiation pressure in the inner accretion disk \citep[e.g.,][]{takeuchi98, gu00, lu04}. Observational evidence for temporal disappearance of the innermost part of the accretion disk, believed to be associated with the \cite{le74} instability, has been found in the black hole X-ray binary {GRS 1915+105} \citep[e.g.,][]{bell97} and V404 Cygni \citep{walton_v404}, which are known to be accreting close to the Eddington limit. In the case of {GRS 1915+105}, the disappearance and follow up replenishment of the inner accretion disk has been reported to be associated with the formation of relativistic expanding clouds and jet ejections \citep[e.g.,][]{pooley97,mira97}.

\subsection{The Spectral Continuum}
Based on the shape of the spectral continuum, the spectral state during the second epoch is similar to the very high/steep power-law state reported in some black hole X-ray binaries, when the luminosity is high and the spectrum is dominated by a steep power-law component \citep[e.g.,][]{bhb_rev06,bell16}. There is evidence that the inner accretion disk in black hole X-ray binaries is sometimes truncated at the very high state. During the very high state, the constant disk emission area inferred from the relation between observed temperature and disc luminosity breaks down \cite[e.g.,][]{kubo04, mccl06}. Detailed modeling of the thermal disk component, taking into account the coupled energetics of the disk and corona, also reveals that the inner disk radius in the very high state is larger than the the location of the ISCO \citep[e.g.,][]{kubo04, done06, tamu12}. By modeling the disk reflection spectra, we have measured a significantly truncated disk in \object\ during the second epoch, supporting this interpretation of the disk structure in the very high state based on previous studies of the thermal disk component. However, because the distance and black hole mass of \object\ are currently unknown, we are unable to obtain an accurate measurement of the inner disk radius via modeling the thermal disk component and directly compare that with the value obtained from the reflection method.

For an approximate estimation, we tried fitting the spectra with the model, {\tt TBabs*simpl*diskbb}, only aiming to describe the shape of the spectral continuum without self-consistently modeling the reflection features. We use the {\tt simpl} convolution model \citep{steiner09} to account for the effect that a fraction of the input disk seed photons are redistributed by Comptonization into a power-law component \citep[e.g.,][]{naka18, shida19}. We only model the disk dominant state spectra, as the simplified  Comptonization model, {\tt simpl}, is not applicable to the power-law dominant state based on the model assumptions \citep{steiner09b}. In order to reveal the intrinsic temperature and luminosity of the thermal disk component during the power-law dominant state, more complicated calculations may be required depending on the disk-corona geometry \citep[e.g.,][]{kubo04, done06,gier08}. The spectral fit yields $kT_{\rm in}\sim1.0$~keV, $N_{\rm diskbb}\sim4000$ for the disk dominant state. Based on the definition of the normalization parameter of {\tt diskbb}, $N_{\rm diskbb}$, and assuming a spectral hardening factor of $f=1.7$ \citep[][]{shimura98}\footnote{$R_{\rm in, ~km}=f^2D_{\rm 10~kpc}\sqrt{N_{\rm diskbb}/{\rm cos}~i}$, where $R_{\rm in}$ is the inner disc radius in km, $f$ is the spectral hardening factor, $D_{\rm 10~kpc}$ is the distance to the source in units of 10 kpc, $N_{\rm diskbb}$ is the normalization of the {\tt diskbb} model, and $i$ is the inner disc inclination. We assume a disk inclination angle of $i=30^{\circ}$ for \object\ when estimating $R_{\rm in}$.}, we estimate the inner disk radius to be $\sim10\times(10~\Msun/M)\times(D/8~{\rm kpc})$~$r_{\rm g}$. We note that there are several possible uncertain factors regarding the thermal disk modeling here. It has been proposed that the spectral hardening factor is likely to be variable \citep[e.g.,][]{mer00,salvesen13, davis19}. Also, at high accretion rates, thermal disk emission deviates from that of an idealized thin accretion disk, which makes the inner disk radius inferred by a thin-disk model problematic at high luminosities \citep[e.g.,][]{mcc06,straub11}.

It is also interesting to consider how the reflection component corresponds to changes of the spectral continuum. Statistical studies of the reflection component in AGNs reveal that disk reflection tends to be less variable than the illuminating continuum \citep[e.g.,][]{mini03, miller07, parker14}. From our observations of \object, we find that the reflection component stays constant across short time intervals (between OBS1 PART I and II, and between OBS2 and OBS3), the variation of the corona emission has only a negligible effect on the shape of the broad iron line. Significant variation in the reflection spectra occurs at the time of state transition, which arises from dynamical changes in the disk structure. However, considering the fact that in several black hole X-ray binaries, previous studies show that the broad iron line remains relatively unchanged in different spectral states \citep[e.g.,][]{reis08, walton12, parker15}, it is possible that the significant change in the shape of the broad iron line we find in \object\ is a peculiar case. 

\subsection{Fast Disk Wind with a Narrow Opening Angle}
Disk winds are believed to be  ubiquitous in X-ray binaries and important to the accretion process, that might trigger instabilities in the accretion flow \citep{begelman83} and even lead to accretion state changes \citep{shields86}. Disk wind launched in X-ray binaries are generally known to have an equatorial geometry, flowing radially at small angles above the accretion disk. Among LMXBs, blueshifted ionized absorption lines in the X-ray spectrum signaling outflowing material are preferentially found in systems viewed close to edge-on \citep[e.g.,][]{ponti12,diaz16}, the majority being ``dippers'' with an inclination estimate of $60^{\circ}-80^{\circ}$ \citep{frank87}.

From the \nustar\ observations of \object, we find a blueshifted ionized absorption feature in the Fe K band in the disk dominant state, which disappears later during the power-law dominant state. The transient nature indicates that \object\ is most likely to be a LMXB. Therefore, the absorption feature found in \object\ probably arises from disk wind rather than the stellar wind from a massive companion star. It is uncommon for strong disk wind features to be detected in a low-inclination black hole binary. We note that the detection of blueshifted ionized iron absorption feature in \object\ (with the disk inclination angle of $i\approx29^{\circ}$, close to face-on) implies that the disk wind observed in \object\ is collimated to be nearly perpendicular to the disk, or has a narrow opening angle assuming a conical geometry. The presence of  disk wind supports that \object\ is accreting at a high accretion rate at the time of our \nustar\ observations. The disappearance of absorption features at the second epoch might be caused by a geometric change of the wind due to the state transition, or over-ionization of the outflowing material due to increased hard X-ray illumination from the central engine \citep[e.g.,][]{miller06, ueda10, diaz14}. It supports a change in the accretion mode between the two epochs. 

In addition, we note that the disk wind velocity we find from the best-fit results is very high, $v_{\rm out}=0.067^{+0.001}_{-0.004}~c$ ($20100^{+300}_{-1200}$~km~s$^{-1}$), significantly exceeding the typical range for disk winds detected in black hole X-ray binaries \citep[e.g.,][]{king13, shida13, diaz16}. This velocity is comparable to some of the most extreme cases that have been reported in black hole X-ray binaries \citep{king12, chiang12, xu_j1658, miller19}, and might be an analogue to the ultra-fast outflows (UFOs) detected in a number of AGNs \citep[e.g.,][]{tombesi2010, tombesi13, nard15}.

\section{Conclusion}
In this work, we report strong and variable relativistic disk reflection features in the new black hole binary candidate \object\ detected by \nustar. Significant difference is found in the strength and the red wing of the broad iron line observed at two epochs, which caught the source in the disk dominant and the power-law dominant state. The iron line flux in the disk dominant state is unusually high when compared with the flux of the weak power-law component, and the iron line flux is not correlated with the strength of the coronal emission. These suggest that the reprocessing of high energy disk photons plays an important role in shaping the disk reflection spectra we observe in the disk dominant state.

By fitting the \nustar\ spectra with self-consistent reflection models, the change in the line profile can be explained by an increase in the inner radius of the optically thick accretion disk: the inner edge of the accretion disk extends down to the ISCO in the disk dominant state and becomes truncated at a larger radius during the power-law dominant state. We discuss the possible physical mechanism (local disk instabilities) that could trigger disk truncation without a significant change in the luminosity/accretion rate. In addition, the results indicate that the central object in \object\ is a rapidly spinning black hole with a spin parameter of $a^*>0.94$, and the accretion disk is viewed at a low inclination angle of $i=29\pm1^{\circ}$. This is also an uncommon case that a disk wind feature is detected in a black hole X-ray binary viewed close to face-on, where the disk wind is found to be extremely fast for a black hole X-ray binary.

\acknowledgments{
We thank the referee for constructive comments that improved the paper. DJW acknowledges support from an STFC Ernest Rutherford fellowship. JAG acknowledges support from NASA grant NNX17AJ65G and from the Alexander von Humboldt Foundation. JH acknowledges support from an appointment to the NASA Postdoctoral Program at the Goddard Space Flight Center, administered by the USRA through a contract with NASA. MLP is supported by European Space Agency (ESA) Research Fellowship. This work was supported under NASA contract No.~NNG08FD60C and made use of data from the \nustar\ mission, a project led by the California Institute of Technology, managed by the Jet Propulsion Laboratory, and funded by the National Aeronautics and Space Administration. We thank the \nustar\ Operations, Software, and Calibration teams for support with the execution and analysis of these observations. This research has made use of the \nustar\ Data Analysis Software (NuSTARDAS), jointly developed by the ASI Science Data Center (ASDC, Italy) and the California Institute of Technology (USA). This research has also made use of MAXI data provided by RIKEN, JAXA and the MAXI team.}

\bibliographystyle{yahapj}
\bibliography{j1631.bib}
\end{document}